# Design of Intelligent layer for flexible querying in databases

Mrs. Neelu Nihalani
Reader, Computer Applications,
UIT, RGPV,
Bhopal, MP India
neelurgpv@yahoo.co.in

Dr. Sanjay Silakari
Prof. & Head, Dept. of CSE,
UIT, RGPV,
Bhopal, MP, India
ssilakari@yahoo.com

Dr. Mahesh Motwani
Reader, Dept. Of CSE
JEC, Jabalpur
Jabalpur, MP, India
mahesh_9@hotmail.com

*Abstract* - **Computer-based information technologies have been extensively used to help many organizations, private companies, and academic and education institutions manage their processes and information systems hereby become their nervous centre. The explosion of massive data sets created by businesses, science and governments necessitates intelligent and more powerful computing paradigms so that users can benefit from this data. Therefore most new-generation database applications demand intelligent information management to enhance efficient interactions between database and the users. Database systems support only a Boolean query model. A selection query on SQL database returns all those tuples that satisfy the conditions in the query.**

**But lately, there is an overwhelming need for non-expert users to query relational databases in their natural language using linguistic variables and terms instead of working with the values of the attributes. As a result, intelligent databases have emerged, which provides expanded and more flexible options for manipulating queries. In this paper, we propose an intelligent layer for database which is responsible for manipulating flexible queries. Initially, the flexible queries from users in their natural language are submitted to intelligent layer and this layer converts the amorphous query into a structured SQL query. The shaped query is executed and the results are presented to the user. Afterwards, on the basis of results, feedback and the acceptance or rejection of the results are requested from the user. It enables the design of a knowledge based self learning system based the values obtained from user, which will aid the selection of appropriate SQL query, when a same flexible query is issued in the future. The experimental results demonstrate the effectiveness of the proposed intelligent database system.**

I.   INTRODUCTION

Databases are gaining prime importance in a huge variety of application areas employing private and public information systems. A general information management system that is capable of managing several kinds of data, stored in the database is known as Database Management System (DBMS) [1]. The DBMS grants support for logical views of data that are separate from the physical views, i.e. how the data is actually stored in the database. By permitting applications to define, access, and update data through a Data Definition Language (DDL) and Data Manipulation Language (DML) combined into a declarative query language such as the relational query language SQL [2] the separation is accomplished.

Structured Query Language (SQL) is an ANSI standard for accessing and manipulating the information stored in relational databases. It is comprehensively employed in industry and is supported by major database management systems (DBMS) [24]. Most of the languages used for manipulating relational database systems are based on the norms of SQL. They work on the basis of Boolean interpretation of the queries: a logical expression is the only accepted selection criterion and the response always encompasses only these tuples for what the expression results in a true value [23]. But some user requirements may not be answered explicitly by a classic querying system. It is due to the fact that the requirements' characteristics cannot be expressed by regular query languages. Many novel-generation database applications stipulate intelligent information management necessitating efficient interactions between the users and database [13]. In recent times, there is a rising demands for non-expert users to query relational databases in a more natural language encompassing linguistic variables and terms, instead of operating on the values of the attributes. Intelligent database systems, a promising approach, enhance the users in performing flexible querying in databases. The research and advancement of intelligent databases have lately emerged as a brand new discipline and have fascinated the attention of numbers.

II. LITERATURE REVIEW

Our work has been inspired by a number of works available in the literature related to intelligent aspects of database systems. The field of intelligent database and information systems has achieved remarkable growth in the last few decades.

Advancements in Intelligent databases focus on two vital issues namely –

- Intelligent information processing in databases.
- Intelligent aspects of databases

*A. Issues of Intelligent information Processing in Databases*

Intelligent information processing has emerged as one of the major considerations on our course to achieve a knowledge society. Recent research in intelligent information processing has paved way for the evolution of thrilling technologies that will mould our future. Intelligent Information Processing is defined as a study on fundamental theory and advanced technology of intelligence and knowledge for information processing. Knowledge-based systems have qualified to offer services in a well-founded ontological framework and there are a number of tools available to support intelligent knowledge management. The techniques of Artificial Intelligence can serve as effective tools in this context. The intelligent systems have a wide range of applications ranging from surfing through the Internet and data-mining, interpreting Internet-derived material, the human Web interface, remote condition monitoring and many other regions [11]. In spite of these applications there are a number of noticeable issues related to intelligent information processing in databases:

A good number of algorithms and approaches on data mining [30, 31], especially association rule mining based algorithms work on the assumption that all items are positively correlated and it considers only those items that remained at last in a shopping basket. Kouris, Makris, and Tsakalidis [32, 33] investigated the mining process by taking valuable information from rejected items and have proposed a number of alternatives for taking the specific items into account efficiently. Another important area of research in Data mining is Outlier detection. Zhao, Bao, Sun, and Yu [12] found the existence of a number of empty cells that are ineffective to outlier detection. They came up with a novel index structure, called CD-Tree which stores only the non empty cells and adopts a clustering technique to store the data objects in the same cell into linked disk pages.

Sampaio et al. [14] proposed an integrated architecture for a Spatial Data warehouse (SDW), including a formalized data model for SDW, a SQL extension query language which facilitates spatial roll-up and drill-down, optimization techniques to improve the performance of complex spatial queries by pre-storing spatial aggregates, and a prototype, Map Warehouse, which validates the ideas proposed. Zarri [15] demonstrated the ubiquity of the "narrative information" and stresses the importance of the same by showing that the traditional ontological tools cannot represent and exploit the narrative information to provide complete and reliable solutions. He also describes the NKRL (Narrative Knowledge Representation Language), a fully-implemented knowledge representation and an inference environment specifically created for an "intelligent" exploitation of narrative knowledge.

The primary goal of most database researches is to incorporate new and related semantics to the data model. Most traditional data models suffer from their inability to manipulate imprecise and vague information that occur in most real world applications. So, we employ the fuzzy set theory in distinct data models and has seen effective solutions for relational and its related models. To comply with modeling of complex objects with imprecision and uncertainty, recent researchers have turned their focus on fuzzy semantic (conceptual) and object-oriented data models. Ma [16] reviewed the fuzzy database modeling technologies, including fuzzy conceptual data models and database models. Regarding fuzzy database models, a brief discussion on fuzzy relational databases and fuzzy object-oriented databases is done.

### B.    Intelligent Aspects of Database

A brief review of some of the works related to intelligent aspects of database systems is illustrated below: Wolff [17] employs the SP theory of computing and cognition to describe some different kinds of "intelligence" exhibited by an intelligent database system. The author introduces the SP theory and its main credits is what that forms the basis for an intelligent database system: that it uses a simple format for diverse kinds of knowledge which is versatile, that it integrates and simplifies a series of AI functions, and that it supports the already established database models when required. The author illustrates the various aspects of "intelligence" in the system: pattern recognition [18] and information retrieval, several forms of probabilistic reasoning, the analysis and production of natural language, and the unsupervised learning of new knowledge, based on a number of examples.

One of the essential characteristic of intelligent database management systems is the ability to provide automated support to users to maintain the semantic correctness of data in compliance to the integrity constraints. These integrity constraints are a vital means to characterize the well-formedness and semantics of the information stored in databases. Martinenghi, Christiansen, and Decker [19] gave an overview of the field of efficient integrity checking and maintenance for relational and deductive databases [20]. The work describes both theoretical and practical aspects of integrity control, including integrity maintenance via active rules. The authors delineate novel scopes of research, particularly with regard to: integrity in XML document [22] collections and in distributed databases, where a strong impact for future developments can be expected. These lines of research pose a number of new and highly relevant research challenges to the database community [21].

III. INTELLIGENT DATABASE SYSTEM (IDBS)

Intelligent database (IDB) systems integrate the resources of both RDBMSs and KBSs to offer a natural way to deal with information, making it easy to store, access and apply [4], [5]. The book "Intelligent Databases" by Kamran Parsaye, Mark Chignell, Setrag Khoshafian and Harry Wong in 1989 was the first to refer the term Intelligent Database [4]. It recommended three levels of intelligence for database systems: High level tools, the user interface and the database engine. The high level tools are meant for administering data quality and unraveling useful relevant patterns automatically by employing a process called Data mining. This layer is highly dependent on the use of artificial intelligence techniques. Intelligent databases encompasses of artificial intelligence components that aids in the intellectual operation of the search, provide means of representing knowledge, and are based on connectionist neural network models [6]. The tasks to be addressed by the intelligent databases are highly complicated, if not impossible, for a human mind to cope with. The tasks involve searching for and extracting meaningful information across a huge data set. . It would be extremely impossible for human minds to deduce, induce or infer any significant new data from the vast data repositories with the efficiency or speed that machine intelligences in the shape of "intelligent" databases achieve [7].

Artificial intelligence is very much able at addressing the difficulties that people are very bad at and perhaps in this context, we consider the "intelligent" databases [7].

Recent researchers in the field of intelligent databases namely Bertino, Catania and Zarri [3] proposed a means to incorporate two technologies explicit; "Intelligent database systems (IDBS) built from the integration of database (DB) technology with techniques developed in the field of artificial intelligence (AI)". Their work also analyzes the inherent weaknesses of the technologies when used in isolation, the traditional databases lacking any semantic value and the inability of artificial intelligence methods to deal with large-data sets. It has been stated by Kamran Parsaye and Mark Chignell [25] that the intelligent databases represent the evolution and merger of several technologies including automatic discovery, Hypermedia, object orientation, expert systems and traditional databases. An intelligent database affords expanded and more flexible options for querying. For example, a user is allowed to type in a question as an imperative sentence. The database then provides a list of hits arranged according to the likelihood (from highest to lowest) that the resulting data contains a useful answer to that question. The Artificial Intelligence (AI) may correct the suspected errors (such as inaccurate spelling) in the input provided by the user [8].

A small number of intelligent databases present synonyms (items with similar meanings) or antonyms (items with opposite or negative meanings) for keywords and phrases. In order to exploit maximum benefit from an intelligent database, the user must formulate queries with forethought, phrasing them with care, just as it is necessary when interrogating a person [8]. As noted in [9], AI/DB integration is crucial for next generation computing, the continued development of

33

DBMS technology and for the effective application of much of AI technology [10]. The motivations driving the integration of these two technologies includes, the need for (a) access to hefty amounts of shared data for knowledge processing, (b) effective management of data as well as knowledge, and (c) intelligent processing of data. In addition to these driving factors, the design of Intelligent Database Interface (IDI) was motivated by the desire to preserve the substantial assets represented by most existing databases. Several general approaches to AI/DB integration have been investigated and stated in the literature [26, 27, 28, 29].

In this research, we propose to develop an intelligent layer which can be incorporated with the existing database system, which is responsible for the intelligent information processing and performing flexible queries. The user queries are given in a more conversing language using linguistic variables and terms. The intelligent layer designed in our scheme processes the unqualified user query and constructs a Standard SQL query from it. Initially, the conjunctive clauses are identified in the user query with the help of the conjunctive training set. Afterwards, on the basis of the identified conjunction, the flexible query is divided into two parts: Subjective/Display part and Criteria part. The subjective/display part contains information about tables and criteria part contains the information about conditions and field names. The expression mapping is carried out in criteria part, which converts expressions into corresponding mathematical symbols. Following expression mapping, the stop words are removed from both the parts of the query. Now, the subjective part contains the table name and the criteria part contains field names and conditions. The next step is to locate the associated tables and fields from the database.

First, we scan through the metadata set for tables to identify the corresponding tables. If the search is unsuccessful, we go for the Ontology based semantic matching, or else the computation of Levenshtein distance, which is a metric employed for measuring the difference between two sequences. The aforesaid procedure is also employed to identify the corresponding field names. From the above data, the SQL query is constructed. The results of the formed SQL query are presented to the user. Subsequently, feedback and the acceptance or rejections of the results are obtained from the user based on the presented results. A knowledge based self learning system is designed with the values obtained from user, which will aid the selection of appropriate SQL query, when a same flexible query is issued in the future. The experimental results demonstrate the effectiveness of the presented intelligent database system. The rest of the paper is organized as follows: A brief review of some of the works in the literature related to intelligent database system is given in Section 2. The proposed intelligent database systems for flexible querying and the knowledge based self learning system are explained in Section 3. The results of the proposed IDBS are presented in Section 4. Finally, the conclusions are summed up in Section 5.

IV. THE PROPOSED INTELIGENT LAYER FOR DATABASES

With advances and in-deep applications of computer technologies, in particular, the extensive applications of Web technology in various areas, databases have become the repositories of large volumes of data.. It is known that databases respond only to standard SQL queries and it is highly impossible for a common person to be well versed in SQL querying. Moreover they may be unaware of the database structures namely table formats, their fields with corresponding types, primary keys and more. On account of these we design an intelligent layer which accepts common user's imperative sentences as input and converts them into standard SQL queries to retrieve data from relational databases based on knowledge base. The primary advantage of the system is that it conceals the inherent complexity involved in information retrieval based on unqualified user queries.

In general, a database ($D$) is termed as set of tables organized in some common structure. The vital information that briefly describes the tables in the database is organized into a metadata set ($M$). The metadata set holds entries for all the '$n$' tables in the relational database with all their corresponding fields and their unique primary key.

The proposed approach employs a set of predefined training structures. The primary benefit of these training sets is that they can be expanded or appended when the intelligent information system discovers some new knowledge. The significant training sets used are: Expression mapping ($E_{map}$), Conjunction set ($C_T$), Semantic set ($S$) for tables and fields are $S_T$ and $S_F$ respectively, Stop-words set ($S_W$).

The Expression mapping set ($E_{map}$) contains the list of commonly used conditional clauses and their associated mathematical symbols. It acts as a look up table to locate the SQL defined mathematical operators. The Conjunction training set ($C_T$) consists of the list of generally used Conjunctive clauses like where, who etc. These conjunctive clauses determine the exact Query definition. When the system encounters a relatively new conjunctive clause, it is appended to the existing training set. The trained stop word set ($S_W$) contains the list of all common stop words that are likely to occur in a user typed query. The semantic set ($S$) contains the list of all possible semantics related to table names and fields in the database.

A. USER QUERY CONTRIBUTION

The following subsection gives a vivid description of how the user query is transformed to be used for data retrieval from databases. In our proposed approach, we define a universal set $Q_u$ which holds all the individual tokens in the user typed query. Every token represents a unique element in the universal set $Q_u$. Every user query is likely to contain a display or subjective part which specifies the intended result, the Conjunction part which determines the SQL definition Clause and the Criteria part which describes the condition or constraint. All these parts of the user query will be represented as three distinct sets $P_d$, $P_j$ and $P_c$. Here the sets $P_d$, $P_j$ and $P_c$ contain tokens that represent the subject, the Conjunctive clause and the Criteria respectively.

$Q_u \rightarrow$ User Query
$P_d \rightarrow$ Subjective/Display part of $Q_u$
$P_j \rightarrow$ Conjunctive part of $Q_u$
$P_c \rightarrow$ Criteria/condition part of $Q_u$

By using the set of predefined training structures, we have determined the Sets $P_d$, $P_j$ and $P_c$ respectively. Then, we analyze the set $P_c$ to locate the conditional clauses provided by the user. The elements of the set $P_c$ are intersected with the trained expression mapping set to map the corresponding mathematical symbols.

Following the expression mapping process, we go for the stop words removal process. This process is meant for eliminating all those words in the user query that are ineffective for SQL query construction. Intersection of the sets $P_d$ and $P_c$ with the trained stop word set $S_w$ will yield a non-empty set of stop words to be eliminated from user query. The above process removes all the stop words from the sets $P_d$ and $P_c$.

Now probably, the set $P_d$ contains the name of the table, the set $P_c$ contains the field name and the condition and the set $P_j$ contains the conjunctive clause from the user query.

B. SUBJECTIVE PART RETRIEVAL

The next step is to locate the table mentioned in the set $P_d$ from $M$. The set $P_d$ is intersected with the metadata set M. If it yields a non-empty set, $P_d$ itself is chosen as the table. Else if it yields an empty set, then the element of singleton set $P_j$ is intersected with the set $S_T$ to retrieve the appropriate table name associated with the matching semantics. Else the distance measure is computed between the element of $P_j$ and every individual element of the set $S_T$. The element in $S_T$ that correlates with minimum Euclidean distance is chosen as the semantics. Finally, the table name associated with the semantics is chosen.

$$S_T \rightarrow \text{Tables in } S$$
$$S_F \rightarrow \text{Fields in } S$$
$$S_{map} \rightarrow S_T \text{ Mapping}$$
$$P_d^t \rightarrow \text{Table } (P_d)$$

Subsequently, we will have to locate the set of related tables which might contain the field name mentioned in the user query. Initially, we scan the set of fields found in $P_d^t$. The procedure is illustrated as below:

$$P_c^f \rightarrow \text{Fields in } P_c$$
$$t_c \rightarrow \text{Table of } P_c^f$$

**Case: 1**

$$(P_c^f \cap f(P_d^t)) \neq \phi$$
$$t_c = P_d^t$$

**Case: 2**

$$t_c = \phi$$

Semantic Map of $\left((P_c^f), S_F(P_d^t)\right)$

*otherwise*

$$t_c = \phi$$

Distance measure of $\left(S_F(P_d^t), (P_c^f)\right)$

If the above procedure results in a non-empty set $t_c$, then to find the related tables, we select the primary key $P_k$ of the table $P_d^t$ from the metadata set and intersect it with all elements in the set $S_F$. If the intersection yields a non-empty set, then the associated tables are appended to a set $t$.

Where $t$ is set of all tables having $K_p$ as a field; where $t \in T$.

Obviously, the number of related tables found for large databases will be enormous. Now to locate a particular field from these related tables is a tedious task, as some of the tables found may be irrelevant to the user query. So, in our proposed approach we introduce a novel step to refine the table set $t$ found. Here, we compute a set $V$ which contains the set of all values corresponding to the field $K_p$ in $t$. If the value set $V$ found for tables in $t$ is a subset of the value set of $K_p$ in $P_d^t$, then append $t$ to $\bar{t}$. Else, eliminate $t_i$.

Subsequently, we go in for the locating the particular field from the set of related tables $\bar{t}$. The same procedure employed to locate the tables is utilized. First, the field

Mrs. Neelu Nihalani et al /International Journal on Computer Science and Engineering Vol.1(2), 2009, 30-39name $F_T$ obtained from the user query is checked in the $M$ followed by the application of semantics and calculation of Euclidean distance.

Finally, we have deduced the appropriate table name $\bar{t}$, field name $t_c$ and the conditional clause $C_q$ from the user query. With all the above information we construct the SQL query to retrieve the requested data from the Database $D$.

C.  SEMANTIC MATCHING

Semantic matching is a technique used to categorize information which is semantically related. Semantic matching is employed as a fundamental technique in many applications areas such as resource discovery, data integration, data migration, query translation, peer to peer networks, agent communication, schema and ontology merging. In fact, it has been accepted as a valid solution to the semantic heterogeneity problem, namely managing the diversity in knowledge. Interoperability among people of different cultures and languages, having different viewpoints and using different terminology has been a massive problem for years. Especially with the advent of the Web and the consequential information explosion, the problem seems to be more emphasized.

D.  DISTANCE MEASURE

The Levenshtein distance is a metric employed for measuring the difference between two sequences (i.e., the so called edit distance). The Levenshtein distance between two strings is defined by the minimum number of operations needed to transform from one string to the other, where an operations can be insertion, deletion, or substitution of a single character. A generalization of the Levenshtein distance (Damerau–Levenshtein distance) permits the transposition of two characters as an operation.

V.  EXPERIMENTAL RESULTS

In this section, we have presented the experimental results of the proposed IDBS. The presented IDBS has been implemented in JAVA with MySQL and MS-Access as databases. The flexible user queries and the results obtained in the designed IDBS are as follows:

A. **Flexible User query:** List orders details where unitprice should be greater than 200
   **Generated SQL Query:** Select * from orders AS A, orderdetails AS B where A.OrderID=B.OrderID and B.UnitPrice > 200

B. **Flexible User query:** List supplier details where city is equal to London.
   **Generated SQL Query:** Select * from suppliers AS A where A. city=London

**Results of query A:**

| OrderID | Custom... | Employ... | OrderD... | Require... | Shipped... | ShipVia | Freight | ShipNa... | ShipAd... | ShipCity | ShipRe... | ShipPo... | ShipCo... | Product | UnitPrice | Quantity | Discount |
|---|---|---|---|---|---|---|---|---|---|---|---|---|---|---|---|---|---|
| 10329 | SPLIR | 4 | 10/15/1... | 11/26/1... | 10/23/1... | 2 | 191.67 | Split Ra... | P.O. Bo... | Lander | WY | 82520 | USA | 10329 | C"te de... | 211 | 20 |
| 10351 | ERNSH | 1 | 11/11/1... | 12/9/19... | 11/20/1... | 1 | 162.33 | Ernst H... | Kirchga... | Graz |  | 8010 | Austria | 10351 | C"te de... | 211 | 20 |
| 10353 | PICCO | 7 | 11/13/1... | 12/11/1... | 11/25/1... | 3 | 360.63 | Piccolo... | Geislw... | Salzburg |  | 5020 | Austria | 10353 | C"te de... | 211 | 50 |
| 10360 | BLONP | 4 | 11/22/1... | 12/20/1... | 12/2/19... | 3 | 131.7 | Blondel... | 24, pla... | Strasbo... |  | 67000 | France | 10360 | C"te de... | 211 | 10 |
| 10372 | QUEEN | 5 | 12/4/19... | 1/1/1997 | 12/9/19... | 2 | 890.78 | Queen... | Alamed... | Sao Pa... | SP | 05487-... | Brazil | 10372 | C"te de... | 211 | 40 |
| 10417 | SIMOB | 4 | 1/16/19... | 2/13/19... | 1/28/19... | 3 | 70.29 | Simons... | Vinb'Ite... | Kobenh... |  | 1734 | Denmark | 10417 | C"te de... | 211 | 50 |

**Results of query B:**

37
ISSN : 0975-3397

| sno | sName | city | status |
|-----|-------|------|--------|
| S1 | Smith | London | 10 |
| S10 | joseph | LONDON | 10 |
| S4 | Steve | London | 10 |
| S5 | George | LONDON | 10 |

From the above pair of examples, we understand that when the SQL query complies with user query the results are expected to be correct.

VI. CONCLUSION

Relational Data Model is universally employed tool for construction of database systems and applications. A dominant technology for data storage and retrieval was developed by Relational Database Management Systems. Yet, these systems struggle the problem of rigidity. Every user requirement cannot be solved by the classic querying system directly because of the requirements' characteristics that are not expressible by standard query languages. In this paper, we have presented an innovative approach for the design of an intelligent database system for performing flexible queries in databases. An intelligent layer has been designed and incorporated into the existing database systems. The presented system accepts flexible user queries and converts them into a standard SQL query. Expression mapping, stop words removal, semantic matching and Levenshtein distance measure techniques have been utilized by the intelligent layer in the formation of the SQL query. The usefulness of the presented system has been demonstrated with the aid of experimental results.